\documentclass[english,conference]{IEEEtran}
\usepackage[T1]{fontenc}
\usepackage[latin9]{inputenc}
\usepackage{verbatim}
\usepackage{float}
\usepackage{amsthm}
\usepackage{amsmath}
\usepackage{amssymb}
\usepackage{graphicx}

\makeatletter

\providecommand{\tabularnewline}{\\}

\theoremstyle{plain}
\newtheorem{thm}{\protect\theoremname}
\theoremstyle{definition}
\newtheorem{defn}[thm]{\protect\definitionname}
\theoremstyle{definition}
\newtheorem{example}[thm]{\protect\examplename}

\usepackage{latexsym}

\usepackage{babel}
\providecommand{\definitionname}{Definition}

\providecommand{\theoremname}{Theorem}

\usepackage{babel}
\providecommand{\definitionname}{Definition}
\providecommand{\examplename}{Example}
\providecommand{\theoremname}{Theorem}

\usepackage{babel}
\providecommand{\definitionname}{Definition}
\providecommand{\examplename}{Example}
\providecommand{\theoremname}{Theorem}

\usepackage{babel}
\providecommand{\definitionname}{Definition}
\providecommand{\examplename}{Example}
\providecommand{\theoremname}{Theorem}

\usepackage{babel}
\providecommand{\definitionname}{Definition}
\providecommand{\examplename}{Example}
\providecommand{\theoremname}{Theorem}

\makeatother

\usepackage{babel}
\providecommand{\definitionname}{Definition}
\providecommand{\examplename}{Example}
\providecommand{\theoremname}{Theorem}

\begin{document}

\title{Non-homogeneous distributed storage systems}

\author{\IEEEauthorblockN{Vo Tam Van, Chau Yuen} \IEEEauthorblockA{Singapore
Univ. of Tech. and Design\\
 Email: \{tamvan\_vo,yuenchau@sutd.edu.sg\}@sutd.edu.sg} \and \IEEEauthorblockN{Jing
(Tiffany) Li} \IEEEauthorblockA{Lehigh University\\
 Email: jingli@lehigh.edu} }

\maketitle

\begin{abstract}
This paper describes a \emph{non-homogeneous} distributed storage
systems (DSS), where there is one super node which has a larger storage
size and higher reliability and availability than the other storage
nodes. We propose three distributed storage schemes based on ($k+2,k$)
maximum distance separable (MDS) codes and non-MDS codes to show the
efficiency of such non-homogeneous DSS in terms of repair efficiency
and data availability. Our schemes achieve optimal bandwidth $\frac{k+1}{2}\frac{M}{k}$
when repairing 1-node failure, but require only one fourth of the
minimum required file size and can operate with a smaller field size
leading to significant complexity reduction than traditional homogeneous
DSS. Moreover, with non-MDS codes, our scheme can achieve an even
smaller repair bandwidth of $\frac{M}{2k}$. Finally, we show that
our schemes can increase the data availability by 10\% than the traditional
homogeneous DSS scheme. \end{abstract}

\section{Introduction}

Distributed storage systems (DSS) are widely used today for storing
data reliably over long periods of time using a distributed collection
of storage nodes, which may be individually unreliable. Application
scenarios include large data centers such as Total Recall \cite{TotalRecall},
OceanStore \cite{Rhea} and peer-to-peer storage systems such as Wuala
\cite{Wuala}, that use nodes across the Internet for distributed
file storage.

One of the challenges for DSS is the \emph{repair problem}: If a node
storing a coded piece fails or leaves the system, in order to maintain
the same level of reliability, we need to create a new encoded piece
and store it at a new node with the minimum repair bandwidth. To solve
this problem, Dimakis \emph{et al.} introduced a generic framework
based on\emph{ regenerating codes }(RC) in \cite{Alex-IT2010}. RC
use ideas from network coding to define a new family of erasure codes
that can achieve different trade-offs in the optimization of storage
capacity and communication costs. The optimal tradeoff curve for achievable
codes has two extremal points which are of particular interest: the
minimum storage regenerating (MSR) codes with minimum possible storage
size for a given repair capability, and the minimum bandwidth regenerating
(MBR) codes with minimum possible repair bandwidth.

Consider a minimum storage system, where a source file of size $M$
units is split into $k$ parts, defined over a finite field $\mathbb{F}_{q}$
and stored across $n$ nodes in the DSS. While the economy in storage
is highly desirable, issues may arise when the system tries to repair
failure at the optimal repair bandwidth. Specifically, if $q$ or
$M$ grows to be arbitrarily large, then the system may become inefficient
and impractical due to the high computational complexity or the fast
growing storage consumption.

Another challenge for DSS is data availability, which is of critical
importance to a peer-to-peer (P2P) storage/backup system that relies
on a swarm of distributed and independent nodes for file storage.
As the nodes not only differ in their storage capacity and traffic
bandwidth, but they may not be online or available at all times. Hence,
there is a pressing need to increase the data availability, such that
infomation is available with a probability approaching 1. Clearly,
P2P enrironments are heterogeneous by nature, and code design for
such systems must explicitly account for this heterogeneity.

The primary interest of this paper is to study a \emph{non-homogeneous
DSS}, where one \emph{\textquotedbl{}super node\textquotedbl{}} has
a larger storage size and higher reliability and availability than
the other storage nodes. We study a class of high-rate $(k+2,k)$
MDS storage codes, and show that with MDS code such non-homogeneous
DSS can achieve the optimal bound in \cite{Alex-IT2010} when repairing
single or double-node failures, but require smaller $M$ and $q$
than the traditional homogeneous model in \cite{Alex-Hadamard}. Another
proposed scheme based on non-MDS codes is shown to repair 1-node failure
below the optimal repair bandwidth bound in \cite{Alex-IT2010}. Moreover,
we show that our proposed non-homogeneous DSS schemes can achieve
a higher data availability than the traditional homogeneous DSS scheme.

This paper is organized as follows. Section II shows the definition
of non-homogeneous DSS. Section III shows three proposed schemes of
exact repair with $(k+2,k)$ storing codes in non-homogeneous DSS.
Section IV shows the numerical results of our schemes and the comparison
with previous methods. Finally, the paper is concluded in Section
V.

\begin{figure}
\begin{centering}
\includegraphics[scale=0.18]{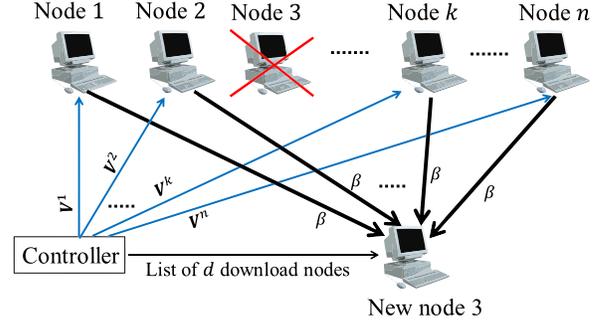} \vspace{-1.5em}
 \caption{An example of traditional repairing 1-node failure.\label{fig:traditionalRepair}}

\par\end{centering}

\centering{}\vspace{-1.5em}

\end{figure}

\section{Models of distributed storage systems\label{sec:definition}}

In this section, we present a brief review of the traditional homogeneous
DSS proposed in \cite{Alex-IT2010}. Then, a new model of non-homogeneous
DSS is proposed to realize the practical DSS.

\subsection{Model of traditional homogeneous DSS}

We follow the definition of traditional homogeneous DSS using $(n,k,d,\alpha,\gamma)$
regenerating codes over finite field $\mathbb{F}_{q}$. This network
has $n$ storage nodes and every $k$ nodes suffice to reconstruct
all the data. The size of the file to be stored is $M$ units%
\footnote{We use ``packets'', ``units'', ``blocks'' interchangeably.%
} and partitioned into $k$ equal parts $\mathbf{f}_{1},\cdots,\mathbf{f}_{k}\in\mathbb{F}_{q}^{N}$
where $N=\frac{M}{k}$. After encoding them into $n$ coded parts
using an $(n,k)$ maximum distance separable (MDS) code, we store
them at $n$ nodes.

We define here the MDS property of a storage code using the notion
of data collectors as presented in \cite{Alex-IT2010}. A storage
code where each node contains $\frac{M}{k}$ worth of storage, has
the MDS property if a data collector can reconstruct the all the $M$
units by connecting to any $k$ out of $n$ storage nodes.

When a node fails, the data stored therein is recovered by downloading
$\beta$ packets each from any $d\;(\geq k)$ of the remaining $(n-1)$
nodes; the total repair bandwidth is then $\gamma=d\beta$ as shown
in Fig. \ref{fig:traditionalRepair}. It has been shown in \cite{Alex-IT2010}
that there exists an optimal tradeoff between the storage per node,
$\alpha$, and the bandwidth to repair one node, $\gamma$. In this
paper, we focus on the extreme point where the smallest $\alpha=\frac{M}{k}$
corresponds to a \emph{minimum-storage regenerating} (\emph{MSR})
code.
\begin{equation}
\left(\alpha_{MSR},\gamma_{MSR}\right)=\left(\frac{M}{k},\frac{Md}{k(d-k+1)}\right)
\end{equation}
 To minimize $\gamma_{MSR}$, let $d=n-1$ and we get $\left(\alpha_{MSR},\gamma_{MSR}^{min}\right)=\left(\frac{M}{k},\frac{M}{k}.\frac{n-1}{n-k}\right)$.
In the case of high-rate codes $n=k+2$, a lower bound for repair
bandwidth $\gamma_{1}$ of 1-node failure was shown as \cite{Alex-IT2010}:
\begin{equation}
\gamma_{1}=(n-1)\beta\geq\frac{M}{k}.\frac{n-1}{n-k}=\frac{M}{k}.\frac{k+1}{2}\label{eq:boundMDS-1}
\end{equation}

\subsection{Model of the proposed non-homogeneous DSS}
\begin{figure}[H]
\centering{}\includegraphics[scale=0.18]{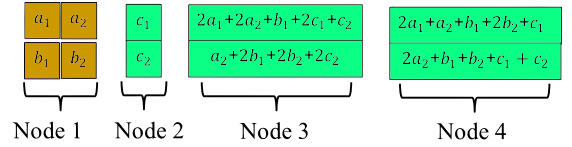} \vspace{-0.5em}
 \caption{An example of non-homogeneous DSS based on $(5,3)$ MDS codes and
$4$ storage nodes.(Node $1$ is the super node.)\label{fig:exampleNon-homogeneousDSS}}
\end{figure}

\begin{defn}
A non-homogeneous DSS with the parameter $(n,k,h)$ is a distributed
storage systems with $h$ nodes based on $(n,k)$ storing codes and
the amount of data stored and downloaded from any nodes are variable.
Node $i$ in the network stores $\alpha_{i}\geq\frac{M}{k}$ units.
When node $i$ fails then it is repaired by downloading $\beta_{j}$
packets from node $j$, $j\in\left\{ n\right\} \backslash i$.\hfill{}$\Box$
\end{defn}
It is clear that we must have $\beta_{j}\leq\alpha_{j}$ for all $j\neq i$
since a node can not transmit more information than it is storing.
When $n=h,\alpha_{i}=\alpha,\beta_{j}=\beta$ for all $i,\; j\neq i$,
we obtain the traditional homogeneous DSS. When $n>h$, there are
more redundant blocks than the storage nodes. The storage process
has to decide which node(s) to store more blocks.
\begin{example}
In this paper, we present the idea of non-homogeneous DSS using the
following setting: there is one big node, called the \emph{super node},
which has a larger storage capacity and higher reliability and availability
than the other nodes. Such scenario is possible in practical system,
e.g. in a peer-to-peer backup system, the super node could be the
service provider that has higher availability and provides higher
storage capacity than other peers.

Consider a system with one super node and three other storage nodes
non-homogeneous DSS based on a $(5,3)$ MDS code, which can be denoted
as $(n=5,k=3,h=4)$. Assume a file of size $M=6$, then this file
is divided into $k=3$ parts, each part containing $N=\frac{M}{k}=2$
packets. After encoding them into $5$ encoded parts or $10$ packets,
we store the first $2N=4$ packets in the super node, and each of
the remaining three nodes stores $N=2$ packets as shown in Fig. \ref{fig:exampleNon-homogeneousDSS}.
\end{example}

\section{Exact repair of $(k+2,k)$ storing codes in non-homogeneous DSS}

In this paper, we limit our study to high-rate $(n=k+2,k)$ exact-repair
storing codes. This homogeneous problem has been considered in \cite{Alex-Hadamard,Cadambe-ISIT,Tamo}.
We propose three efficient DSS schemes using MDS and non-MDS storage
codes in such $(n=k+2,k,h=k+1)$ non-homogeneous DSS, which are denoted
as Scheme A, B, and C in Table. \ref{table:(k+2,k)model}. Scheme
A and C use MDS codes while Scheme B uses non-MDS codes. The new system
consists of $(k+1)$ nodes which include $k$ nodes of storage size
$N$ and one super node of size $2N$.

\begin{table*}[t]
\centering \caption{Three schemes of $(k+2,k,k+1)$ non-homogeneous model vs. traditional
model based on $(k+2,k)$ MDS codes where S and P are the abbreviation
of systematic and parity, respectively. Here, $\mathbf{f}_{i}\in\mathbb{F}_{q}^{1\times N}$
and $\mathbf{A}_{i},\mathbf{B}_{i}\in\mathbb{F}_{q}^{N\times N}$
for all $1\leq i\leq k$. Note that all Schemes A, B and C use only
$(k+1)$ storage nodes to store $(k+2)$ packets. \label{table:(k+2,k)model}}

\begin{tabular}{|c|c|c|c|}
\hline
 & \multicolumn{2}{c|}{Non-homogeneous} & Homogeneous\tabularnewline
\hline
 & Proposed Scheme A and B  & Proposed Scheme C  & Traditional model \cite{Alex-Hadamard}\tabularnewline
\hline
S. node  &  &  & \tabularnewline
\hline
$s_{1}$  & $\begin{array}{c}
\mathbf{f}_{1\quad}\mathbf{f}_{2}\end{array}$  & $\mathbf{f}_{1}$  & $\mathbf{f}_{1}$\tabularnewline
\hline
$s_{2}$  & $\mathbf{f}_{3}$  & $\mathbf{f}_{2}$  & $\mathbf{f}_{2}$\tabularnewline
\hline
$\vdots$  & $\vdots$  & $\vdots$  & $\vdots$\tabularnewline
\hline
$s_{k-1}$  & $\mathbf{f}_{k}$  & $\mathbf{f}_{k-1}$  & $\mathbf{f}_{k-1}$\tabularnewline
\hline
$s_{k}$  & $\times$  & $\mathbf{f}_{k}$  & $\mathbf{f}_{k}$\tabularnewline
\hline
P. node  &  &  & \tabularnewline
\hline
$p_{1}$  & $\mathbf{f}_{1}\mathbf{A}_{1}+\cdots+\mathbf{f}_{k}\mathbf{A}_{k}$  & $\begin{array}{c}
\mathbf{f}_{1}\mathbf{A}_{1}+\cdots+\mathbf{f}_{k}\mathbf{A}_{k}\\
\mathbf{f}_{1}\mathbf{B}_{1}+\cdots+\mathbf{f}_{k}\mathbf{B}_{k}
\end{array}$  & $\mathbf{f}_{1}\mathbf{A}_{1}+\cdots+\mathbf{f}_{k}\mathbf{A}_{k}$\tabularnewline
\hline
$p_{2}$  & $\mathbf{f}_{1}\mathbf{B}_{1}+\cdots+\mathbf{f}_{k}\mathbf{B}_{k}$  & $\times$  & $\mathbf{f}_{1}\mathbf{B}_{1}+\cdots+\mathbf{f}_{k}\mathbf{B}_{k}$\tabularnewline
\hline
\end{tabular}
\end{table*}

\subsection{Scheme A: Store two systematic data at the super node with MDS codes. }

\begin{figure}
\centering{}\includegraphics[scale=0.16]{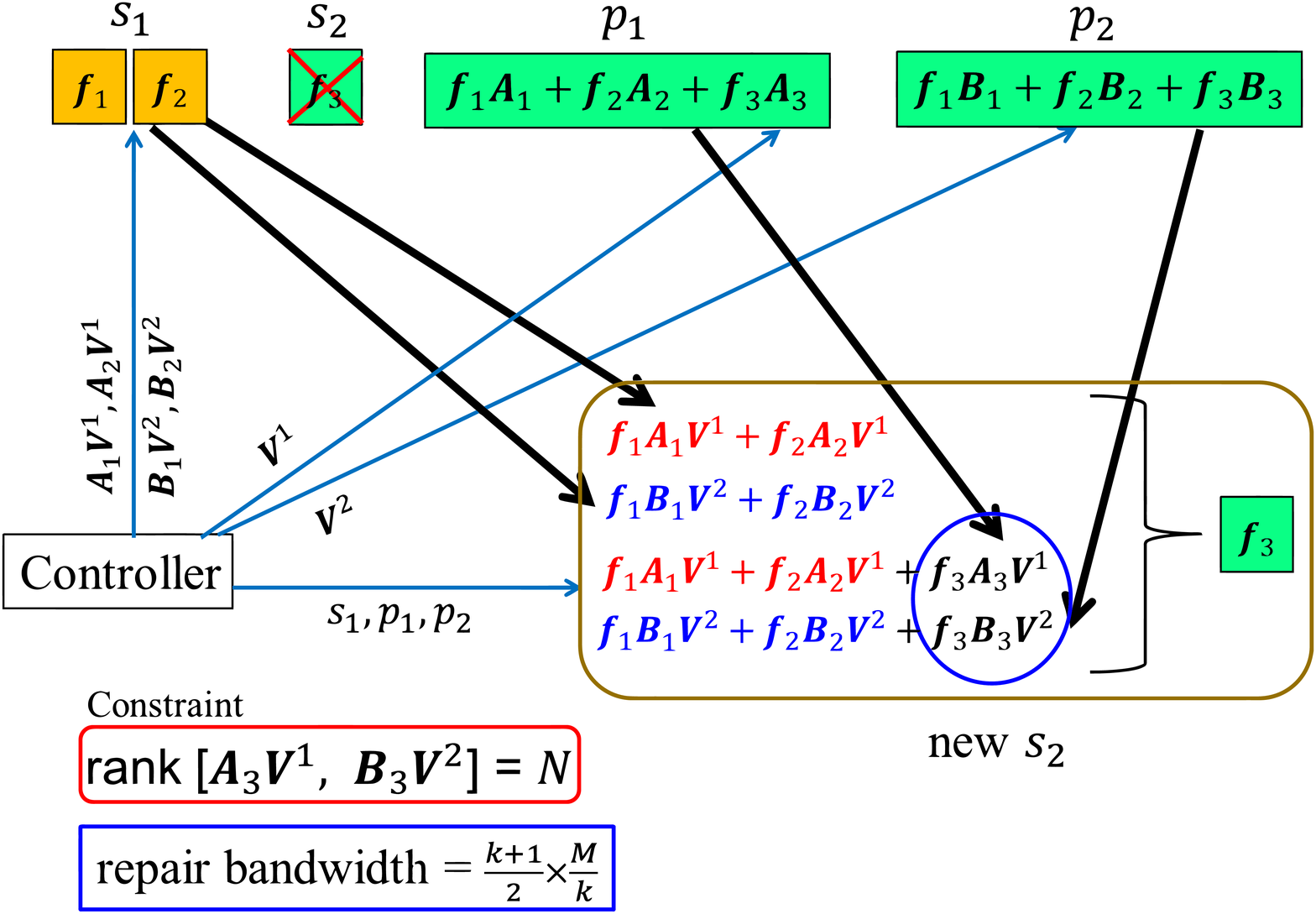} \vspace{-0.5em}
 \caption{An example of repairing 1-node failure using Scheme A based on $(5,3)$
MDS codes in non-homogeneous DSS\label{fig:1repairSchemeAn5k3}}
\end{figure}

It can be seen in Table. \ref{table:(k+2,k)model}, the first $(k-1)$
storage nodes of Scheme A store the systematic file parts $\mathbf{f}_{1},\cdots,\mathbf{f}_{k}$
where $\mathbf{f}_{1},\mathbf{f}_{2}$ are stored in the same systematic
node $s_{1}$ and the other file parts $\mathbf{f}_{3},\cdots,\mathbf{f}_{k}$
are stored individually in the remaining $(k-2)$ systematic nodes
$s_{2},\cdots,s_{k-1}$, respectively. The first and second parity
$p_{1}$, $p_{2}$ store a linear combination of all $k$ systematic
parts as $\mathbf{f}_{1}\mathbf{A}_{1}+\cdots+\mathbf{f}_{k}\mathbf{A}_{k}$
and $\mathbf{f}_{1}\mathbf{B}_{1}+\cdots+\mathbf{f}_{k}\mathbf{B}_{k}$.
Here, $\mathbf{A}_{i}$ and $\mathbf{B}_{i}$ denote an $N\times N$
matrix of coding coefficients defined over finite field $\mathbb{F}_{q}$.
\begin{itemize}
\item \textbf{Repair systematic failure nodes for Scheme A}:
\end{itemize}
We first consider repairing 1-node failure (in the case of super node
$s_{1}$ fails, it is considered as 2-node failures which will be
discussed in more detail later). Without loss of generality we assume
node $s_{2}$ that contains $\mathbf{f}_{3}$ is failed. For simplicity,
we first consider the case $(n=5,k=3$). To recover desired data $\mathbf{f}_{3}$,
we have to download the following equations from the two survival
parity nodes:

\begin{equation}
\left\{ \hspace{-0.5em}\begin{array}{c}
\mathbf{f}_{1}\mathbf{A}_{1}\mathbf{V}^{1}+\mathbf{f}_{2}\mathbf{A}_{2}\mathbf{V}^{1}+\mathbf{f}_{3}\mathbf{A}_{3}\mathbf{V}^{1}\\
\mathbf{f}_{1}\mathbf{B}_{1}\mathbf{V}^{2}+\mathbf{f}_{2}\mathbf{B}_{2}\mathbf{V}^{2}+\mathbf{f}_{3}\mathbf{B}_{3}\mathbf{V}^{2}
\end{array}\right.
\end{equation}
 where $\mathbf{A}_{i},\mathbf{\, B}_{i}\in\mathbb{F}_{q}^{N\times N}$
for all $1\leq i\leq k$ and $\mathbf{V}^{1},\mathbf{V}^{2}\in\mathbb{F}_{q}^{N\times\frac{N}{2}}$
can be derived based on the failure node. To repair different failure
nodes, different $\mathbf{V}^{1},\mathbf{V}^{2}$ are needed which
can be precalculated. It can be seen from Fig. \ref{fig:1repairSchemeAn5k3}
that the term $\left(\mathbf{f_{\textrm{1}}}\mathbf{A}_{1}\mathbf{V}^{1}+\mathbf{f_{\mathrm{2}}}\mathbf{A}_{2}\mathbf{V}^{1}\right)$
and $\left(\mathbf{f_{\textrm{1}}}\mathbf{B}_{1}\mathbf{V}^{2}+\mathbf{f_{\textrm{2}}}\mathbf{B}_{2}\mathbf{V}^{2}\right)$
are removable by downloading $\left(\frac{N}{2}+\frac{N}{2}\right)$
packets from super node. Therefore, the desired data $\mathbf{f}_{3}$
can be recovered if the following rank constraint is satisfied:

\begin{equation}
\begin{array}{c}
\mathrm{rank}\left[\mathbf{A}_{3}\mathbf{V}^{1},\;\mathbf{B}_{3}\mathbf{V}^{2}\right]=N\end{array}\label{eq:rankSchemeA}
\end{equation}

To recover the desired data $\mathbf{f}_{3}$ in the general $(n=k+2,k)$
case, we have to use the following equations:

\begin{equation}
\left\{ \hspace{-0.5em}\begin{array}{c}
\mathbf{f_{\textrm{1}}}\mathbf{A}_{1}\mathbf{V}^{1}+\mathbf{f}_{2}\mathbf{A}_{2}\mathbf{V}^{1}+\mathbf{f}_{3}\mathbf{A}_{3}\mathbf{V}^{1}+\cdots+\mathbf{f}_{k}\mathbf{A}_{k}\mathbf{V}^{1}\\
\mathbf{f_{\textrm{1}}}\mathbf{B}_{1}\mathbf{V}^{2}+\mathbf{f}_{2}\mathbf{B}_{2}\mathbf{V}^{2}+\mathbf{f}_{3}\mathbf{B}_{3}\mathbf{V}^{2}+\cdots+\mathbf{f}_{k}\mathbf{B}_{k}\mathbf{V}^{2}
\end{array}\right.\label{eq:termA1B1}
\end{equation}

Similarly, the term $\left(\mathbf{f_{\textrm{1}}}\mathbf{A}_{1}\mathbf{V}^{1}+\mathbf{f_{\mathrm{2}}}\mathbf{A}_{2}\mathbf{V}^{1}\right)$
and $\left(\mathbf{f_{\textrm{1}}}\mathbf{B}_{1}\mathbf{V}^{2}+\mathbf{f}_{2}\mathbf{B}_{2}\mathbf{V}^{2}\right)$
are removable by downloading $\left(\frac{N}{2}+\frac{N}{2}\right)$
packets from super node 1. The following conditions must be satisfied
to achieve the optimal repair bandwidth:

\begin{equation}
\begin{array}{c}
\mathrm{rank}\left[\mathbf{A}_{3}\mathbf{V}^{1},\;\mathbf{B}_{3}\mathbf{V}^{2}\right]=N\\
\mathrm{rank}\left[\mathbf{A}_{4}\mathbf{V}^{1},\;\mathbf{B}_{4}\mathbf{V}^{2}\right]=\frac{N}{2}\\
\vdots\\
\mathrm{rank}\left[\mathbf{A}_{k}\mathbf{V}^{1},\;\mathbf{B}_{k}\mathbf{V}^{2}\right]=\frac{N}{2}
\end{array}\label{eq:condition1repair-2systematicAB}
\end{equation}

To relax the complexity of the constraints found in (\ref{eq:condition1repair-2systematicAB}),
we set $\mathbf{A}_{i}=\mathbf{I}_{N}$ and $\mathbf{V}^{1}=\mathbf{V}^{2}$,
then obtain the following equations:

\begin{equation}
\begin{array}{c}
\mathrm{rank}\left[\mathbf{B}_{3}\mathbf{V}^{1},\;\mathbf{V}^{1}\right]=N\\
\mathrm{rank}\left[\mathbf{B}_{4}\mathbf{V}^{1},\;\mathbf{V}^{1}\right]=\frac{N}{2}\\
\vdots\\
\mathrm{rank}\left[\mathbf{B}_{k}\mathbf{V}^{1},\;\mathbf{V}^{1}\right]=\frac{N}{2}
\end{array}\label{eq:condition1repair-2systematic-1}
\end{equation}

The problem of finding matrix $\mathbf{B}_{i}$ is similar to \cite{Alex-Hadamard}.
However, here we only need to solve $(k-2)$ equations. Therefore,
the fragment size and finite field will be smaller, with $M=2^{k-1}k$
(which means the fragment size reduce to $\frac{1}{4}$ of the traditional
homogeneous model when $k\geq3$), and $q=2k-1$. These advantages
allow us to reduce the minimum size unit of storing file and reduce
the complexity of computation to a smaller finite field. It can be
seen that in this case of 1-node failure, the proposed Scheme A can
achieve the optimal repair bandwidth of $\frac{k+1}{2}\frac{M}{k}$,
which is the same as traditional homogenous DSS scheme.

\begin{itemize}
\item \textbf{Repair first parity node for Scheme A}:
\end{itemize}
If the first parity node $p_{1}$ fails, we make a change of variable
to obtain a new representation for our code such that the first parity
$p_{1}$ becomes a systematic node in the new representation. We make
the change of variables as follows:

\begin{equation}
\sum_{i=1}^{k}\mathbf{f}_{i}=\mathbf{y}_{3},\;\mathbf{f}_{s}=\mathbf{y}_{s}\quad\mathrm{for}\;1\leq s\neq3\leq k\label{eq:firstParitySchemeA}
\end{equation}
 We solve (\ref{eq:firstParitySchemeA}) by replacing $\mathbf{f}_{3}$
in terms of the $\mathbf{y}_{i}$ variables and obtain
\[
\mathbf{f}_{3}=\mathbf{y}_{3}-\left(\mathbf{y}_{1}+\mathbf{y}_{2}+\mathbf{y}_{4}+\cdots+\mathbf{y}_{k}\right)
\]

The problem of repairing first parity is equivalent to repair systematic
node $\mathbf{y}_{3}$ in the new presentation. Note that $\mathbf{y}_{1},\mathbf{y}_{2}$
are stored in the same node since they are correspondent to $\mathbf{f}_{1},\mathbf{f}_{2}$.
To repair $\mathbf{y}_{3}$, we have to download the following equations
from node $s_{2}$ and $p_{2}$:

\[
\left\{ \hspace{-0.5em}\begin{array}{l}
(-\mathbf{y}_{1})+(-\mathbf{y}_{2})+\mathbf{y}_{3}+\cdots+(-\mathbf{y}_{k})\\
(\mathbf{B}_{1}-\mathbf{B}_{3})\mathbf{y}_{1}+(\mathbf{B}_{2}-\mathbf{B}_{3})\mathbf{y}_{2}+\mathbf{B}_{3}\mathbf{y}_{3}+\cdots+(\mathbf{B}_{k}-\mathbf{B}_{3})\mathbf{y}_{k}
\end{array}\right.
\]

Again, the $\mathbf{V}^{1},\mathbf{V}^{2}$ matrices need to satisfy
the following conditions in order to achieve the optimal repair bandwidth.

\begin{equation}
\begin{array}{l}
\mathrm{rank}\left[\mathbf{B}_{3}\mathbf{V}^{1},\;\mathbf{V}^{1}\right]=N\\
\mathrm{rank}\left[(\mathbf{B}_{4}-\mathbf{B}_{3})\mathbf{V}^{1},\;\mathbf{V}^{1}\right]=\frac{N}{2}\\
\qquad\qquad\qquad\vdots\\
\mathrm{rank}\left[(\mathbf{B}_{k}-\mathbf{B}_{3})\mathbf{V}^{1},\;\mathbf{V}^{1}\right]=\frac{N}{2}
\end{array}\label{eq:condition1repair-2systematic-paritynode}
\end{equation}

\begin{figure}
\begin{centering}
\includegraphics[scale=0.14]{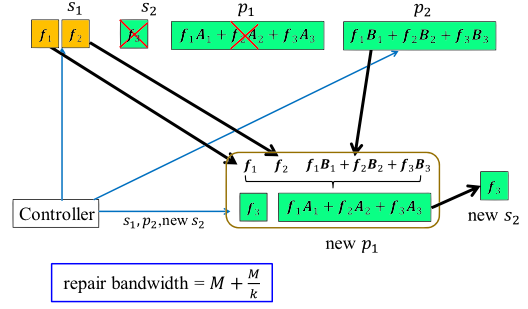}
\par\end{centering}

\vspace{0.5em}

\centering{}\caption{Illustration of exact repair of 2-node failure for a (5,3) Exact-Repair
MDS code for Scheme A in non-homogeneous DSS. Total repair bandwidth
$\gamma_{2}=M+\frac{M}{k}$ achieves lower bandwidth bound.\label{fig:graph2repairOptimal}}
\end{figure}

Similar to the systematic case, the solution of matrix $\mathbf{B}_{i}$
is similar to \cite{Alex-Hadamard}. However, we need to solve only
$(k-2)$ equations, which means the fragment size and the finite field
will be smaller $M=2^{k-1}k$, and $q=2k-1$.
\begin{itemize}
\item \textbf{Repair second parity node for Scheme A}:
\end{itemize}
Similar to the above, we rewrite this code in a form where the second
parity is a systematic node in some presentation

\begin{equation}
\begin{array}{l}
\left[\begin{array}{ccccc}
\mathbf{I}_{N} & 0 & 0 & \cdots & 0\\
0 & \mathbf{I}_{N} & 0 & \cdots & 0\\
0 & 0 & \mathbf{I}_{N} & \cdots & 0\\
\vdots & \vdots & \vdots & \vdots & \vdots\\
0 & 0 & 0 & \cdots & \mathbf{I}_{N}\\
\mathbf{I}_{N} & \mathbf{I}_{N} & \mathbf{I}_{N} & \cdots & \mathbf{I}_{N}\\
\mathbf{B}_{1} & \mathbf{B}_{2} & \mathbf{B}_{3} & \cdots & \mathbf{B}_{k}
\end{array}\right]\mathbf{f}\\
\hspace{8em}=\left[\begin{array}{ccccc}
\mathbf{I}_{N} & 0 & 0 & \cdots & 0\\
0 & \mathbf{I}_{N} & 0 & \cdots & 0\\
0 & 0 & \mathbf{I}_{N} & \cdots & 0\\
\vdots & \vdots & \vdots & \vdots & \vdots\\
0 & 0 & 0 & \cdots & \mathbf{I}_{N}\\
\mathbf{B}_{1}^{'} & \mathbf{B}_{2}^{'} & \mathbf{B}_{3}^{'} & \cdots & \mathbf{B}_{k}^{'}\\
\mathbf{I}_{N} & \mathbf{I}_{N} & \mathbf{I}_{N} & \cdots & \mathbf{I}_{N}
\end{array}\right]\mathbf{f}^{'}
\end{array}\label{eq:secondParityEquation}
\end{equation}
 where $\mathbf{f}^{'}$ is a full rank row transformation of $\mathbf{f}$.
We proceed in a way similar to how we handled the first parity repair
to achieve the optimal repair bandwidth. A similar set of equations
to the case of repairing the first parity node can be obtained as
shown below.

\begin{equation}
\begin{array}{l}
\mathrm{rank}\left[\mathbf{B}_{3}^{'}\mathbf{V}^{1},\;\mathbf{V}^{1}\right]=N\\
\mathrm{rank}\left[(\mathbf{B}_{4}^{'}-\mathbf{B}_{3}^{'})\mathbf{V}^{1},\;\mathbf{V}^{1}\right]=\frac{N}{2}\\
\qquad\qquad\qquad\vdots\\
\mathrm{rank}\left[(\mathbf{B}_{k}^{'}-\mathbf{B}_{3}^{'})\mathbf{V}^{1},\;\mathbf{V}^{1}\right]=\frac{N}{2}
\end{array},\label{eq:condition1repair-2parity}
\end{equation}
 It can be seen that in this case the size and the finite field will
be again $M=2^{k-1}k$, and $q=2k-1$, which are smaller than those
in \cite{Alex-Hadamard}, and still achieve the optimal repair bandwidth.
\begin{itemize}
\item \textbf{Repair 2-node failures for Scheme A}
\end{itemize}
To repair 2-node failure at the optimal repair bandwidth, one solution
is shown in Fig. \ref{fig:graph2repairOptimal}. Lets assume that
$s_{2}$ and $p_{1}$ fail, to repair them, first download the $k$
packets from the survival nodes, then the original file can be recovered
due to the property of MDS codes. Therefore, we can obtain the data
of node $s_{2}$ and $p_{1}$, and store them in new node, say new
$p_{1}$. Next, the data of the failure node $s_{2}$ (i.e. $\mathbf{f_{3}}$
in this case) is forwarded to the new node $s_{2}$. The total repair
bandwidth will be $\gamma_{2}=M+\frac{M}{k}$. It is trivial to repair
the super node $s_{1}$ at the repair bandwidth of $M$ by downloading
data from survival nodes. It should be noted that the failure of one
super node plus one additional node cannot be repaired since it can
be regarded as three-node failure, therefore beyond the correcting
ability of $(k+2,k)$ MDS codes.

\subsection{Scheme B: Store two systematic data at the super node with non-MDS
codes. }

Scheme B uses the same model as Scheme A. However, we can achieve
the repair bandwidth of 1-node failure below the optimal bound in
this non-homogeneous model if the term $\left(\mathbf{f_{\textrm{1}}}\mathbf{A}_{1}\mathbf{V}^{1}+\mathbf{f_{\mathrm{2}}}\mathbf{A}_{2}\mathbf{V}^{1}\right)$
and $\left(\mathbf{f_{\textrm{1}}}\mathbf{B}_{1}\mathbf{V}^{2}+\mathbf{f}_{2}\mathbf{B}_{2}\mathbf{V}^{2}\right)$
in (\ref{eq:termA1B1}) are the same or the following constraints
are satisfied $\mathbf{A}_{1}\mathbf{V}^{1}=\mathbf{\lambda B}_{1}\mathbf{V}^{2},\mathbf{A}_{2}\mathbf{V}^{1}=\lambda\mathbf{B}_{2}\mathbf{V}^{2}$.
It means that we only need to download $\frac{N}{2}$ packets instead
of $\left(\frac{N}{2}+\frac{N}{2}\right)$ packets from the super
node to eliminate these terms. The following example is used to present
the idea of repairing 1-node failure below the optimal bandwidth bound
for the case $k=3,n=5$. Suppose $\mathbf{f}_{1}=[a_{1},a_{2}]^{T},\mathbf{f}_{2}=[b_{1},b_{2}]^{T},\mathbf{f}_{3}=[c_{1},c_{2}]^{T}$
and $\mathbf{p}_{1}=\mathbf{f}_{1}\mathbf{A}_{1}+\mathbf{f}_{2}\mathbf{A}_{2}+\mathbf{f}_{3}\mathbf{A}_{3},$
$\mathbf{p}_{2}=\mathbf{f}_{1}\mathbf{B}_{1}+\mathbf{f}_{2}\mathbf{B}_{2}+\mathbf{f}_{3}\mathbf{B}_{3}$
are the systematic and parity data of a $(5,3)$ storage code over
finite field $\mathbb{F}_{3}$ where

\begin{equation}
\begin{array}{c}
\mathbf{A}_{1}=\left[\begin{array}{cc}
2 & 0\\
2 & 1
\end{array}\right],\;\mathbf{A}_{2}=\left[\begin{array}{cc}
1 & 2\\
0 & 2
\end{array}\right],\;\mathbf{A}_{3}=\left[\begin{array}{cc}
2 & 0\\
1 & 2
\end{array}\right],\\
\mathbf{B}_{1}=\left[\begin{array}{cc}
2 & 0\\
1 & 2
\end{array}\right],\;\mathbf{B}{}_{2}=\left[\begin{array}{cc}
1 & 1\\
2 & 1
\end{array}\right],\;\mathbf{B}_{3}=\left[\begin{array}{cc}
1 & 1\\
0 & 1
\end{array}\right].
\end{array}
\end{equation}

\begin{figure}
\centering{}\includegraphics[scale=0.16]{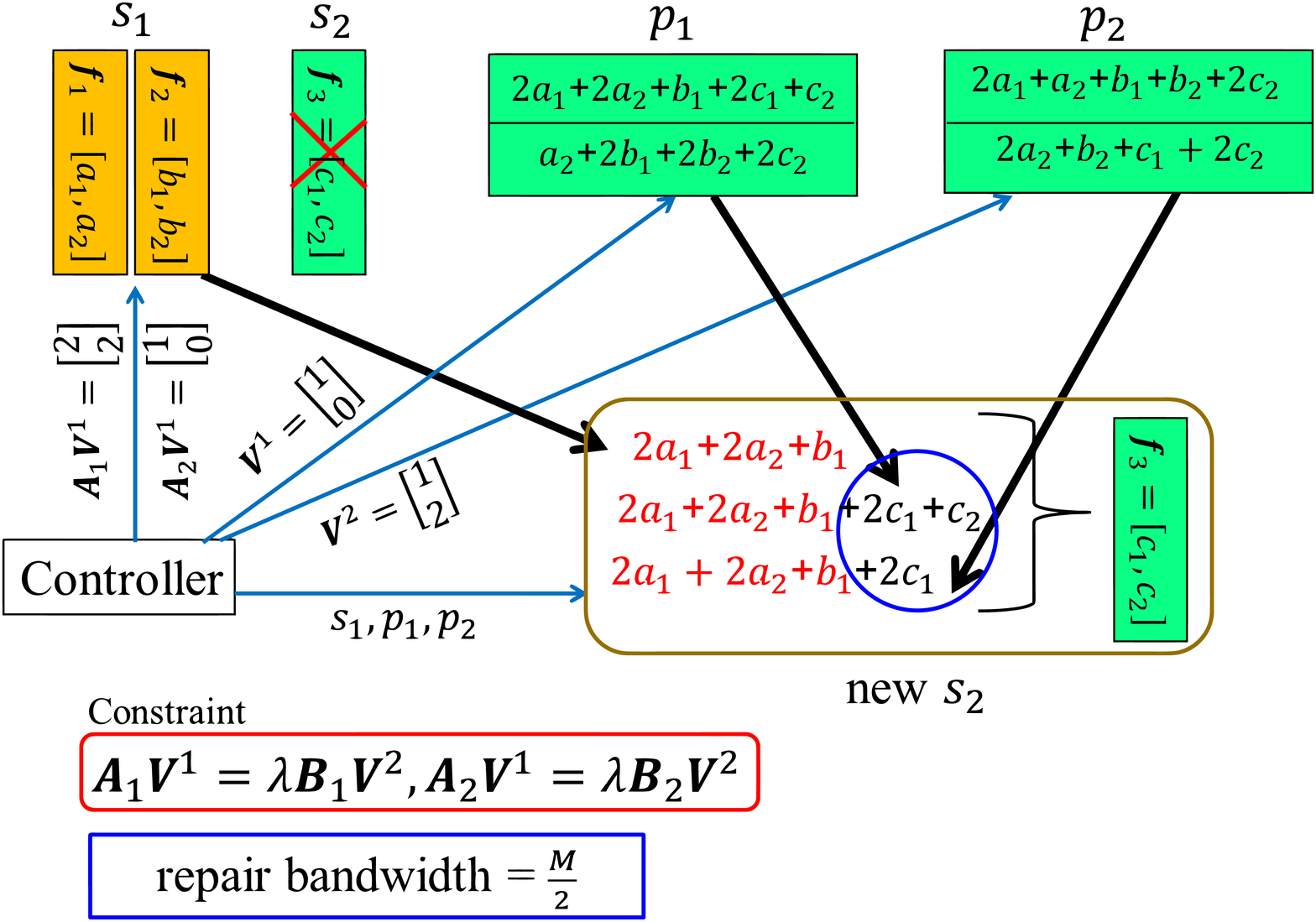} \vspace{-0.5em}
 \caption{Total repair bandwidth of 1-node failure $\gamma_{1}=\frac{M}{2}$
is smaller than the bound. In this example, $\gamma_{1}=3<4$ of repair
bandwidth in the traditional case\label{fig:graph1repairBeyondBound-2systematicnode-1}}
\end{figure}

It can be seen that any 1-node failure (systematic or parity node)
except the super node can be repaired with bandwidth of $\frac{M}{2}$,
which is below the optimal bound $\left(\frac{k+1}{2}\frac{M}{k}\right)$.
Fig \ref{fig:graph1repairBeyondBound-2systematicnode-1} shows the
process of using two projection vectors $\mathbf{V}_{1}=\left[\begin{array}{c}
1\\
0
\end{array}\right],\mathbf{V}_{2}=\left[\begin{array}{c}
1\\
2
\end{array}\right]$ for repairing one systematic node failure below the optimal bandwidth
bound. It is straightforward for the general case $(k+2,k)$. In the
case of systematic 1-node failure, the general design constraints
for Scheme B is:

\begin{equation}
\begin{array}{c}
\mathbf{A}_{1}\mathbf{V}^{1}=\mathbf{\lambda B}_{1}\mathbf{V}^{2}\\
\mathbf{A}_{2}\mathbf{V}^{1}=\lambda\mathbf{B}_{2}\mathbf{V}^{2}\\
\mathrm{rank}\left[\mathbf{A}_{3}\mathbf{V}^{1},\;\mathbf{B}_{3}\mathbf{V}^{2}\right]=N\\
\mathrm{rank}\left[\mathbf{A}_{4}\mathbf{V}^{1},\;\mathbf{B}_{4}\mathbf{V}^{2}\right]=\frac{N}{2}\\
\vdots\\
\mathrm{rank}\left[\mathbf{A}_{k}\mathbf{V}^{1},\;\mathbf{B}_{k}\mathbf{V}^{2}\right]=\frac{N}{2}
\end{array}\label{eq:condition1repair-nonMDS}
\end{equation}

Similar to Scheme A, we can find the solution for scheme B. However,
in Scheme B the MDS property of the storage code breaks since we cannot
reconstruct the original information from the survival nodes in the
case of 2-node failure.

\subsection{Scheme C: Store two parity data at the super node with MDS codes. }

\begin{figure}
\centering{}\includegraphics[scale=0.16]{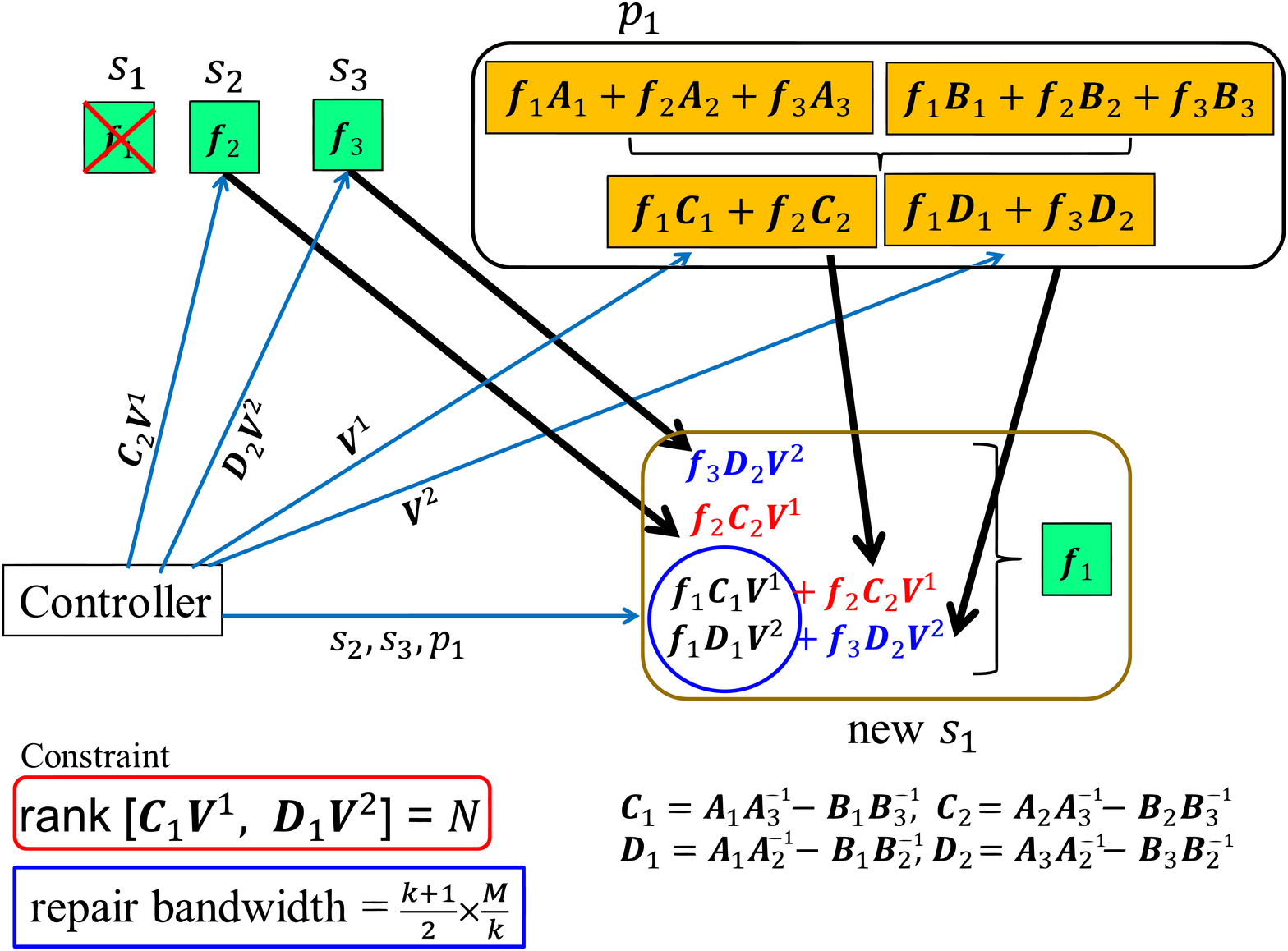} \vspace{-0.5em}
 \caption{An example of repairing 1-node failure using Scheme C based on $(5,3)$
MDS codes in non-homogeneous DSS\label{fig:1repairSchemeBn5k3}}
\end{figure}

We first consider an example with $n=5,k=3$ for simplicity. Without
loss of generality, assume that node $1$ with data $\mathbf{f}_{1}$
is failed and the two parity packets $\mathbf{p}_{1},\mathbf{p}_{2}$
are stored at the super node. To recover $\mathbf{f}_{1}$, we have
the following equations after eliminating $\mathbf{f}_{2}$ and $\mathbf{f}_{3}$
from the parity node:

\begin{equation}
\left\{ \hspace{-0.5em}\begin{array}{c}
\mathbf{\mathbf{f}_{\mathrm{1}}\mathbf{A}_{\mathrm{1}}V^{\mathrm{1}}}+\mathbf{f}_{\mathrm{2}}\mathbf{\mathbf{A}_{\mathrm{2}}V^{\mathrm{1}}+\mathbf{f}_{\mathrm{3}}A_{\mathrm{3}}V^{\mathrm{1}}}\\
\mathbf{\mathbf{f}_{\mathrm{1}}\mathbf{B}_{\mathrm{1}}V^{\mathrm{2}}}+\mathbf{\mathbf{f}_{\mathrm{2}}B_{\mathrm{2}}V^{\mathrm{2}}+\mathbf{f}_{\mathrm{3}}\mathbf{B}_{\mathrm{3}}V^{\mathrm{2}}}
\end{array}\right.\hspace{-0.5em}\rightarrow\left\{ \hspace{-0.5em}\begin{array}{c}
\mathbf{\mathbf{f}_{\mathrm{1}}\mathbf{C}_{\mathrm{1}}V^{\mathrm{1}}}+\mathbf{\mathbf{f}_{\mathrm{2}}\mathbf{C}_{\mathrm{2}}V^{\mathrm{1}}}\\
\mathbf{\mathbf{f}_{\mathrm{1}}\mathbf{D}_{\mathrm{1}}V^{\mathrm{2}}}+\mathbf{f}_{\mathrm{3}}\mathbf{\mathbf{D}_{\mathrm{2}}V^{\mathrm{2}}}
\end{array}\right.
\end{equation}
 where $\mathbf{C}_{i},\mathbf{D}_{i}\in\mathbb{F}_{q}^{N\times N}$
for $i=1,2$ and $\mathbf{C}_{1}=\mathbf{A}_{1}\mathbf{A}_{3}^{-1}-\mathbf{B}_{1}\mathbf{B}_{3}^{-1},\mathbf{C}_{2}=\mathbf{A}_{2}\mathbf{A}_{3}^{-1}-\mathbf{B}_{2}\mathbf{B}_{3}^{-1},\mathbf{D}_{1}=\mathbf{A}_{1}\mathbf{A}_{2}^{-1}-\mathbf{B}_{1}\mathbf{B}_{2}^{-1},\mathbf{D}_{2}=\mathbf{A}_{3}\mathbf{A}_{2}^{-1}-\mathbf{B}_{3}\mathbf{B}_{2}^{-1}$.
It can be seen from Fig. \ref{fig:1repairSchemeBn5k3} that the term
$\mathbf{f_{\mathrm{\mathrm{\textrm{2}}}}\mathbf{C}_{\mathrm{2}}V^{\mathrm{1}}}$
and $\mathbf{f}_{\mathrm{\textrm{3}}}\mathbf{D}_{\mathrm{2}}\mathbf{V}^{\mathrm{2}}$
are removable by downloading $\left(\frac{N}{2}+\frac{N}{2}\right)$
packets from the parity node. Therefore, the desired data $\mathbf{f}_{1}$
can be recovered if the following rank constraint is satisfied:

\begin{equation}
\begin{array}{c}
\mathrm{rank}\left[\mathbf{C}_{1}\mathbf{V}^{1},\;\mathbf{D}_{1}\mathbf{V}^{2}\right]=N\end{array}\label{eq:rankSchemeB}
\end{equation}

For general $(k+2,k)$ case, similar to Scheme A, we set $\mathbf{A}_{i}=\mathbf{I}_{N}$
for all $i\leq N$. To recover the desired data $\mathbf{f}_{1}$,
we have the following equations reduction from parity node:

\begin{equation}
\left\{ \hspace{-0.5em}\begin{array}{c}
\mathbf{f}_{1}\left(\mathbf{B}_{1}-\mathbf{B}_{3}\right)+\mathbf{f}_{2}\left(\mathbf{B}_{2}-\mathbf{B}_{3}\right)+\sum_{j=4}^{k}\mathbf{f}_{j}\left(\mathbf{B}_{j}-\mathbf{B}_{3}\right)\\
\mathbf{f}_{1}\left(\mathbf{B}_{1}-\mathbf{B}_{2}\right)+\mathbf{f}_{3}\left(\mathbf{B}_{3}-\mathbf{B}_{2}\right)+\sum_{j=4}^{k}\mathbf{f}_{j}\left(\mathbf{B}_{j}-\mathbf{B}_{2}\right)
\end{array}\right.
\end{equation}

The following conditions must be satisfied to achieve the optimal
repair bandwidth of $\frac{k+1}{2}\frac{M}{k}$:

\begin{equation}
\begin{array}{c}
\mathrm{rank}\left[\left(\mathbf{B}_{1}-\mathbf{B}_{2}\right)\mathbf{V}^{1},\;\left(\mathbf{B}_{1}-\mathbf{B}_{3}\right)\mathbf{V}^{2}\right]=N\\
\mathrm{rank}\left[\left(\mathbf{B}_{4}-\mathbf{B}_{2}\right)\mathbf{V}^{1},\;\left(\mathbf{B}_{4}-\mathbf{B}_{3}\right)\mathbf{V}^{2}\right]=\frac{N}{2}\\
\vdots\\
\mathrm{rank}\left[\left(\mathbf{B}_{k}-\mathbf{B}_{2}\right)\mathbf{V}^{1},\;\left(\mathbf{B}_{k}-\mathbf{B}_{3}\right)\mathbf{V}^{2}\right]=\frac{N}{2}
\end{array}\label{eq:condition1repair}
\end{equation}

In general, solving (\ref{eq:condition1repair}) is still an open
problem. Here, we give a numerical solution for the case $n=6,k=4$.
Consider $\mathbf{f}_{1}=[a_{1},a_{2}]^{T},\mathbf{f}_{2}=[b_{1},b_{2}]^{T},\mathbf{f}_{3}=[c_{1},c_{2}]^{T},\mathbf{f}_{4}=[c_{1},c_{2}]^{T}$
and $\mathbf{p}_{1}=\mathbf{f}_{1}+\mathbf{f}_{2}+\mathbf{f}_{3}+\mathbf{f}_{4},$
$\mathbf{p}_{2}=\mathbf{f}_{1}\mathbf{B}_{1}+\mathbf{f}_{2}\mathbf{B}_{2}+\mathbf{f}_{3}\mathbf{B}_{3}+\mathbf{f}_{4}\mathbf{B}_{4}$
are the systematic and parity data of a $(6,4)$ storage code over
finite field $\mathbb{F}_{3}$ where

\begin{equation}
\begin{array}{c}
\mathbf{B}_{1}=\left[\begin{array}{cc}
0 & 1\\
1 & 0
\end{array}\right],\;\mathbf{B}_{2}=\left[\begin{array}{cc}
0 & 2\\
2 & 0
\end{array}\right],\\
\mathbf{B}_{3}=\left[\begin{array}{cc}
2 & 0\\
0 & 1
\end{array}\right],\;\mathbf{B}_{4}=\left[\begin{array}{cc}
1 & 1\\
1 & 2
\end{array}\right].
\end{array}
\end{equation}

\begin{figure}
\centering{}\includegraphics[scale=0.16]{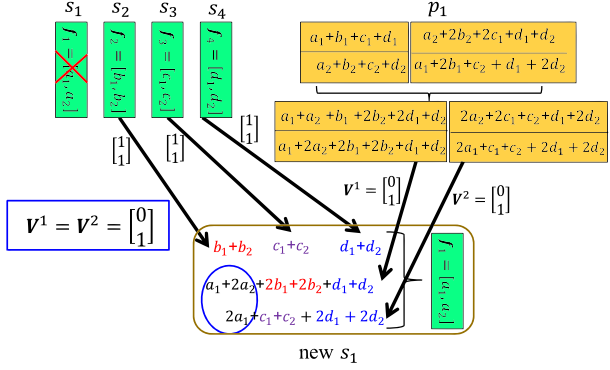} \vspace{-1.5em}
 \caption{An example of repairing 1-node failure using Scheme C based on $(6,4)$
MDS codes in non-homogeneous DSS\label{fig:graph1repairSchemeC_N6K4}}
\end{figure}

It can be seen that any 1-node failure except the super node can be
repaired with optimal bandwidth $\left(\frac{k+1}{2}\frac{M}{k}\right)$.
Fig. \ref{fig:graph1repairSchemeC_N6K4} shows the process of using
two projection vectors $\mathbf{V}^{1}=\mathbf{V}^{2}=\left[\begin{array}{c}
0\\
1
\end{array}\right]$ for repairing the first systematic node.

For the case of super node $p_{1}$ and 2-node fail, the repair process
is similar to scheme A. The total repair bandwidth will be $M$ and
$M+\frac{M}{k}$, respectively.\textbf{ }

\section{Performance Analysis}

As compared with the previous work in \cite{Alex-Hadamard}, our schemes
A and C can achieve optimal repair bandwidth of 1-node failure at
a smaller finite field $q$ and 75\% smaller data size $M$ than \cite{Alex-Hadamard}.
Moreover, Scheme B that uses non-MDS code can repair 1-node failure
with $\frac{M}{2k}$ smaller bandwidth than the optimal bound. A summary
is presented in Table \ref{table:comparison} for the various technologies.

\begin{table*}
\centering \caption{Comparison of non-homogeneous model vs. traditional model based on
$(k+2,k)$ codes where $M$ and $\gamma$ are the file size and repair
bandwidth, respectively. Small value of $M,\gamma$ and $q$ mean
efficient.\label{table:comparison}}

\vspace{0.5em}
\begin{tabular}{|c|c|c|c|c|c|c|}
\hline
 & Scheme A\&C  & Scheme B  & Alex \cite{Alex-Hadamard}  & Perm. code \cite{Cadambe-ISIT}  & Tamo \cite{Tamo}  & C.R.C \cite{Kenneth}\tabularnewline
\hline
 & $\begin{array}{c}
M=2^{k-1}k\\
q\geq2k-1
\end{array}$  & $\begin{array}{c}
M=2^{k-1}k\\
q\geq2k-1
\end{array}$  & $\begin{array}{c}
M=2^{k+1}k\\
q\geq2k+3
\end{array}$  & $\begin{array}{c}
M=2^{k}k\\
q\geq2k+1
\end{array}$  & $\begin{array}{c}
M=2^{k}k\\
q\geq2k+1
\end{array}$  & $\begin{array}{c}
M=2k\\
q\geq n
\end{array}$\tabularnewline
\hline
$1$- node failure  & $\gamma=\frac{M}{k}\frac{k+1}{2}$  & $\gamma=\frac{M}{2}$  & $\gamma=\frac{M}{k}\frac{k+1}{2}$  & $\gamma=\frac{M}{k}\frac{k+1}{2}$  & $\gamma=\frac{M}{k}\frac{k+1}{2}$  & N.A\tabularnewline
\hline
$2$- node failures  & $\begin{array}{c}
\gamma=M+\frac{M}{k}\end{array}$  & N.A  & $\begin{array}{c}
\gamma=M+\frac{M}{k}\end{array}$  & $\begin{array}{c}
\gamma=M+\frac{M}{k}\end{array}$  & $\begin{array}{c}
\gamma=M+\frac{M}{k}\end{array}$  & $\begin{array}{c}
\gamma=M+\frac{M}{k}\end{array}$\tabularnewline
\hline
\end{tabular}
\end{table*}

\subsection{Numerical Case Study}

To show an illustration, we continue using the example $n=5,\, k=3$.
Assume we have the data file of size $48$ blocks to store across
the DSS.

\emph{Scheme A} \emph{and C}: Divide the file into 4 fragments of
size $M_{1}=12$ blocks. These fragments are stored across $k+1=4$
nodes in the non-homogeneous DSS. In the case of 1-node failure, the
repair bandwidth will be $4\times\frac{M_{1}}{k}\frac{k+1}{2}=32$
blocks. In the case of 2-node failure, the repair bandwidth will be
$4\times\left(M_{1}+\frac{M_{1}}{k}\right)=64$ blocks. To update
one fragment $M_{1}$ of the file, the update bandwidth will be $\frac{M_{1}}{k}n=20$
blocks.

\emph{Scheme B}: Similar to Scheme A, the file is divided into 4 fragments
of size $M_{1}=12$. In the case of 1-node failure, the repair bandwidth
will be $4\times\frac{M_{1}}{2}=24$. To update one fragment $M_{1}$
of the file, the update bandwidth will be $\frac{M_{1}}{k}n=20$.

\emph{Alex method} \cite{Alex-Hadamard}: Divide the file into 1 fragment
of size $M_{2}=48$. The repair bandwidth of 1-node failure will be
$1\times\left(\frac{M_{2}}{k}\frac{k+1}{2}\right)=32$ and for 2-node
failure requires $1\times\left(M_{2}+\frac{M_{2}}{k}\right)=64$.
The update bandwidth of one fragment $M_{2}$ will be $\frac{M_{2}}{k}n=80$.
The repair and update bandwidth of other methods are computed in the
same manner and shown as in Table. \ref{table:numericalResult}. Note
that C.R.C method cannot repair 1-node failure with optimal bandwidth.

From Table. \ref{table:numericalResult}, it can be seen that Schemes
A and C can achieve the optimal bandwidth for repairing 1- or 2-node
failure, which is the same as homogeneous DSS. By scarifying the MDS
property, Scheme B requires a lower repair bandwidth for 1-node failure.
It can be seen that all proposed schemes have an advantage of small
update bandwidth in compare with the other schemes except the CRC
method. However, the CRC method is not practical since it cannot achieve
the optimal repair bandwidth in the case of 1-node failure. \cite{Cadambe-ISIT}
and \cite{Tamo} methods are also impractical since they can repair
only the systematic nodes.

\begin{table*}[t]
\centering \caption{Numerical results of storing a file of size 48 blocks using the $(5,3)$
MDS codes in non-homogeneous and homogeneous DSS.\label{table:numericalResult}}

\vspace{0.5em}
\begin{tabular}{|c|c|c|c|c|c|c|}
\hline
 & Scheme A\&C  & Scheme B  & Alex\cite{Alex-Hadamard}  & Perm. code\cite{Cadambe-ISIT}  & Tamo \cite{Tamo}  & C.R.C \cite{Kenneth}\tabularnewline
\hline
\hline
 & $\begin{array}{c}
M_{1}=12\\
q\geq5
\end{array}$  & $\begin{array}{c}
M_{1}=12\\
q\geq5
\end{array}$  & $\begin{array}{c}
M_{2}=48\\
q\geq9
\end{array}$  & $\begin{array}{c}
M_{3}=24\\
q\geq7
\end{array}$  & $\begin{array}{c}
M_{4}=24\\
q\geq7
\end{array}$  & $\begin{array}{c}
M_{5}=6\\
q\geq5
\end{array}$\tabularnewline
\hline
$1$-node failure  & $\begin{array}{c}
\gamma=32\end{array}$  & $\begin{array}{c}
\gamma=24\end{array}$  & $\begin{array}{c}
\gamma=32\end{array}$  & $\begin{array}{c}
\gamma=32\end{array}$  & $\begin{array}{c}
\gamma=32\end{array}$  & N.A\tabularnewline
\hline
$2$-node failures  & $\gamma=64$  & N.A  & $\gamma=64$  & $\gamma=64$  & $\gamma=64$  & $\gamma=64$\tabularnewline
\hline
$\begin{array}{c}
\mathrm{update}\\
\leq12\mathrm{\; data}\;\mathrm{block}
\end{array}$  & $\delta=20$  & $\delta=20$  & $\delta=80$  & $\delta=40$  & $\delta=40$  & $\delta=10$\tabularnewline
\hline
\end{tabular}
\end{table*}

\subsection{Data Availability}

In this subsection, we employ the framework proposed in \cite{HeterogeneousDSS}
to measure and compare the data availability between our proposed
non-homogenous DSS schemes and traditional homogenous DSS schemes
to show the efficiency of our proposed schemes. Let $\left[p_{1},\cdots,p_{h}\right]$
be the nodes' online probability of $h$ nodes in the $(n,k,h)$ DSS.
Let the power set of $h$, $2^{h}$, denote the set of all possible
combinations of online nodes. Let $A\subset2^{h}$ represents one
of these possible combinations. Then, we will use $Q_{A}$ to represent
the event that combination $A$ occurs. Since node availabilities
are independent, we have

\begin{equation}
Pr[Q_{A}]=\prod_{i\in A}p_{i}\prod_{j\in2^{h}\backslash A}\left(1-p_{j}\right)
\end{equation}
 Let $x_{i}$ be the number of data blocks stored in storage node
$i$, for example $x_{i}=1$, it means $\alpha_{i}=\frac{M}{k}$.
The data allocation of our schemes will be $\left(x_{1}=2,x_{2}=1,\cdots,x_{k+1}=1,x_{k+2}=0\right)$.
Let $L_{k}\subset2^{h}$ be the subset containing those combinations
of available nodes which together store $k$ different redundant blocks.
\begin{equation}
L_{k}=\left\{ A:\; A\in2^{h},\sum_{i\in A}x_{i}\geq k\right\}
\end{equation}
 Since the retrieval process needs to download $k$ different blocks
out of the total $n$ redundant blocks, the probability of successful
recovery for an allocation $\left(x_{1},\cdots,x_{n}\right)$ can
be measured as
\begin{equation}
\begin{array}{c}
Pr\left[\mathrm{successful\; recovery}\right]=\sum_{A\in L_{k}}Pr\left[Q_{A}\right]\\
=\sum_{A\in L_{k}}\left[\prod_{i\in A}p_{i}\prod_{j\in2^{h}\backslash A}\left(1-p_{j}\right)\right]
\end{array}
\end{equation}
 To compare the data availability, we examine a scenario of node online
probability where the online probability of super node is greater
than the other node $p_{1}\geq p_{2}=p_{3}=\cdots=p_{n}=p$. The data
availability of homogeneous $Pr_{homo}$ (e.g. scheme in \cite{Alex-Hadamard})
and non-homogeneous $Pr_{non-homo}$ DSS (e.g. Schemes A and C, since
Scheme B is based on non-MDS code, it is excluded in this study as
its availability is calculated in a different manner) can be computed
by the following equations:

\begin{equation}
Pr_{homo}=p^{k+1}+(k+1)(1-p)p^{k}+\frac{k(k+1)}{2}p_{1}(1-p)^{2}p^{k-1}
\end{equation}

\begin{equation}
Pr_{non-homo}=p^{k}+kp_{1}(1-p)p^{k-1}+\frac{k(k+1)}{2}p_{1}(1-p)^{2}p^{k-1}
\end{equation}

\begin{figure}
\centering{}\includegraphics[scale=0.35]{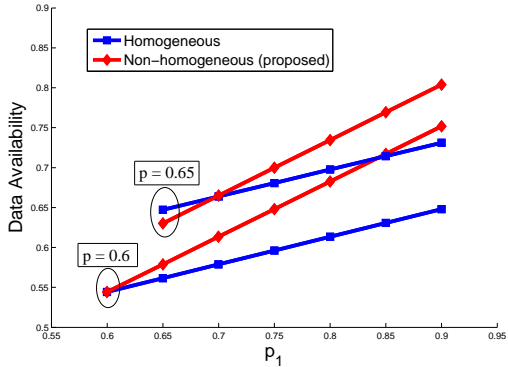} \vspace{-0.5em}
 \caption{A comparision of data availability between non-homogeneous DSS and
homogeneous DSS.\label{fig:dataAvailable}}
\end{figure}

Let $p_{1}=\chi p$ where $\chi\geq1$. The condition $Pr_{non-homo}\geq Pr_{homo}$
will induce $\chi\geq p/(p+\frac{1}{2}(1-p)\left[(k-1)-(k+1)p\right])$.
It can be seen that if $p\leq\frac{k-1}{k+1}$, then $p/(p+\frac{1}{2}(1-p)\left[(k-1)-(k+1)p\right])\leq1\leq\chi$.
Therefore, $Pr_{non-homo}\geq Pr_{homo}$ for all $p\leq\frac{k-1}{k+1}$.
We run the simulations for the case of $k=4,\: p=0.6$ and $p=0.65$
and obtain the result in Fig. \ref{fig:dataAvailable}. It can be
seen that for $p=\frac{k-1}{k+1}=0.6$, data availability of non-homogeneous
DSS scheme outperforms the homogeneous DSS scheme. For $p=0.65>\frac{k-1}{k+1}$,
the non-homogeneous schemes also have a big improvement when $p_{1}$
has a high online availability. Therefore, it can be seen that our
proposed non-homogeneous DSS schemes achieve a higher data availability
than the traditional homogeneous DSS. The gap between the two becomes
larger when the online availability of the super node increases, e.g.
when $p_{1}$ is greater than 25\% of $p$, the data availability
of the proposed non-homogenous over homogenous DSS is increased by
10\%.

\section{Conclusions}

We proposed three distributed storage schemes for \emph{non-homogeneous
DSS} with high rate $(k+2,k)$ codes. Two of the schemes make use
of MDS code, and can achieve optimal repair bandwidth of $\frac{k+1}{2}\frac{M}{k}$
at smaller finite field $q$ and 75\% smaller fragment $M$ than \cite{Alex-IT2010}.
Small $M$ and $q$ are desirable, because they reduce the update
bandwidth and complexity. Another scheme based on non-MDS code can
achieve a smaller repair bandwidth than the optimal bandwidth based
on MDS code by $\frac{M}{2k}$ for 1-node failure. We further demonstrate
that in such non-homogeneous DSS, if we can ensure one super node
with a higher online probability than the other nodes, we can achieve
a higher data availability than the homogeneous DSS.

\section*{Acknowledgement}

This research is partly supported by the International Design Center
(grant no. IDG31100102 \& IDD11100101). Li's work is supported in
part by the National Science Foundation under Grants No.
CCF-0829888, CMMI-0928092, and EAGER-1133027.


\begin{thebibliography}{10}
\bibitem{Alex-Hadamard} D.S. Papailiopoulos, A.G. Dimakis, V.R. Cadambe,
``Repair optimal erasure codes through Hadamard designs,'' \emph{the
49th Annual Allerton Conference on Communication, Control and Computation},
pp.1382-1389, Sep. 2011.

\bibitem{Alex-IT2010} A.G. Dimakis, P. Godfrey, M. Wainwright, and
K. Ramchandran, ``Network coding for distributed storage systems,\textquotedblright{}
\emph{IEEE Trans. Inform. Theory}, pp.4539-4551, Sep. 2010.

\bibitem{Alex-ISIT2009} Y.Wu, A.G. Dimakis, ``Reducing repair traffic
for erasure coding based storage via interference alignment,'' \emph{IEEE
International Symposium on Information Theory}, pp.2276-2280, July
2009.

\bibitem{TotalRecall} R. Bhagwan, K. Tati, Y.-C. Cheng, S. Savage,
and G. M. Voelker, ``Total recall: System support for automated availability
management,\textquotedblright{}\emph{ the 1st Symp. Networked Systems
Design and Implementation (NSDI)}, Mar. 2004.

\bibitem{Cadambe-ISIT} V.R. Cadambe, C. Huang, J. Li, ``Permutation
code: optimal exact-repair of a single failed node in MDS code based
distributed storage systems,'' \emph{IEEE International Symposium
on Information Theory,} pp.1225-1229, Aug. 2011.

\bibitem{Cadambe-Asilomar} V.R. Cadambe, S.A. Jafar, H. Maleki, ``Asymptotic
interference alignment for exact repair in distributed storage systems,''
\emph{the 44th Signals, Systems and Computers (ASILOMAR)}, pp.1617-1621,
Nov. 2010.

\bibitem{Cullina} D. Cullina, A. G. Dimakis, and T. Ho, ``Searching
for minimum storage regenerating codes,'' \emph{Allerton Conf. Control
Comput. Commun.}, Sep. 2009.

\bibitem{DHT}F. Dabek, J. Li, E. Sit, J. Robertson, M. Kaashoek,
and R. Morris, ``Designing a DHT for low latency and high throughput,\textquotedblright{}
\emph{the 1st} \emph{Symp. Networked Systems Design and Implementation
(NSDI)}, Mar. 2004.

\bibitem{Wuala}D. Grolimund, ``Wuala - A Distributed File System\textquotedblright{},
Google Tech Talk http://www.youtube.com/watch?v=3xKZ4KGkQY8

\bibitem{Kenneth}Cooperative regenerating codes, http://home.ie.cuhk.edu.hk/\textasciitilde{}wkshum/
papers/CRC4.pdf

\bibitem{Kumar} N.B. Shah, K.V. Rashmi, P.V. Kumar, ``A flexible
class of generating codes for distributed storage,'' \emph{IEEE International
Symposium on Information Theory}, pp.1943-1947, Jun. 2010.

\bibitem{Rhea}S. Rhea, C. Wells, P. Eaton, D. Geels, B. Zhao, H.
Weatherspoon, and J. Kubiatowicz, ``Maintenance-free global data
storage,\textquotedblright{} \emph{IEEE Internet Computer}, pp.40-49,
Sep. 2011.

\bibitem{Suh} C. Suh, K. Ramchandran, ``Exact-repair MDS code construction
using interference alignment,'' \emph{IEEE Trans. Inform. Theory},
pp.1425-1442, Mar. 2011.

\bibitem{Tamo} I. Tamo, Z. Wang, J. Bruck, ``MDS array codes with
optimal rebuilding,'' \emph{IEEE International Symposium on Information
Theory}, pp.1240-1244, Aug. 2011.

\bibitem{Vardy} C.Armstrong, A. Vardy, ``Distributed storage with
communication cost,'' \emph{the 49th Annual Allerton Conference on
Communication, Control and Computation}, pp.1358-1365, Sep. 2011.

\bibitem{HeterogeneousDSS} L. Pamies-Juarez, P. Garcia-Lopez, M.
Sanchez-Artigas, B. Herrera, ``Towards the design of optimal data
redundancy schemes for heterogeneous cloud storage infrastructures'',
Computer Network, pp. 1100-1113, Nov. 2011. \end{thebibliography}
\end{document}